\documentclass[doublecol]{epl2}
\usepackage{graphicx}
\usepackage{amsmath}
\usepackage{amssymb}
\usepackage{tabularx}

\title{Klein paradox for a pn junction in multilayer graphene}

\author{B. Van Duppen\thanks{E-mail: \email{ben.vanduppen@ua.ac.be}} \and F. M. Peeters\thanks{E-mail: \email{francois.peeters@ua.ac.be}}}

\institute{Department of Physics, University of Antwerp, Groenenborgerlaan 171, B-2020 Antwerp, Belgium}


\abstract{Charge carriers in single and multilayered graphene systems behave as chiral particles due to the particular lattice symmetry of the crystal. We show that the interplay between the meta-material properties of graphene multilayers and the pseudospinorial properties of the charge carriers result in the occurrence of Klein and anti-Klein tunneling for rhombohedral stacked multilayers. We derive an algebraic formula predicting the angles at which these phenomena occur and support this with numerical calculations for systems up to four layers. We present a decomposition of an arbitrarily stacked multilayer into pseudospin doublets that have the same properties as rhombohedral systems with a lower number of layers.}

\begin{document}

\maketitle

\section{Introduction}

The electronic properties of the one atom thick graphene crystal has been the subject of several recent papers\cite{Neto2009, Rozhkov2011} since its experimental isolation.\cite{Novoselov2004} Not only is the crystal a promising candidate for semiconductor physics, the electron behavior also mimics that of Dirac particles and can therefore be seen as an interesting table top realization of a two dimensional quantum relativistic system. The Klein paradox, a unit transmission probability through potential barriers of any height or width, was one of the first characterizing phenomena from QED predicted\cite{Katsnelson2006} and subsequently observed experimentally\cite{Stander2009}. But also optical properties as Fabry-P\'{e}rot resonances\cite{RamezaniMasir2010} and the negative refraction index that makes graphene a metamaterial\cite{Cheianov2007} are remarkable properties of the crystal. The stacking of two layers of graphene, the so called bilayer graphene (BLG), although being only weakly bound, changes fundamentally the electronic properties. For example, the Klein paradox as observed in monolayer graphene (MLG) is replaced by the suppression of transmission which is called anti-Klein tunneling\cite{Campos}. This suppressed transmission is remarkable since there are hole states available inside a potential barrier that are cloaked from the continuum of states outside\cite{Gu2011}. For trilayer graphene (TLG), Klein tunneling is present if the layers are orthorhombic stacked. For Bernal stacking however Klein tunneling is absent\cite{Kumar2012,VanDuppen2012}.

The occurrence of these tunneling phenomena is the consequence of the lattice induced chiral nature of the charge carriers. When the Fermi energy of the electrons is low enough, in MLG the dispersion of the electrons in the vicinity of one of the Dirac points \textbf{K} is linear in reciprocal space. This allows for the introduction of pseudospin, which is the lattice induced analogue of conventional spin from the Dirac theory. Extending this concept to BLG, the dispersion near the Dirac point can be approximated as being parabolic but due to symmetry arguments still leads to a spinorial Hamiltonian describing the electrons as chiral particles with a pseudospin\cite{McCann2006, VanDuppen2012b}.

Some recent papers have discussed the electronic structure\cite{Min2008, Koshino} and known concepts such as trigonal warping and the Berry phase\cite{Morimoto2012,Mikitik2008,Koshino2009a} of graphene multilayers. In this paper we generalize the discussions of Katsnelson \textit{et al.}\cite{Katsnelson2006} and Gu \textit{et al.}\cite{Gu2011} to an arbitrary number of layers and to an arbitrary stacking order for a pn junction. We find that the low energy behaviour can be expressed as a system of non interacting pseudospin doublets, each with a specific chirality and derive a simple algebraic expression for the angles at which Klein tunneling (KT) and anti-Klein tunneling (AKT) can be expected.

\section{Structure of the multilayer}

\label{Structure}

\begin{figure}[tb]
\begin{center}
\includegraphics[width=8cm]{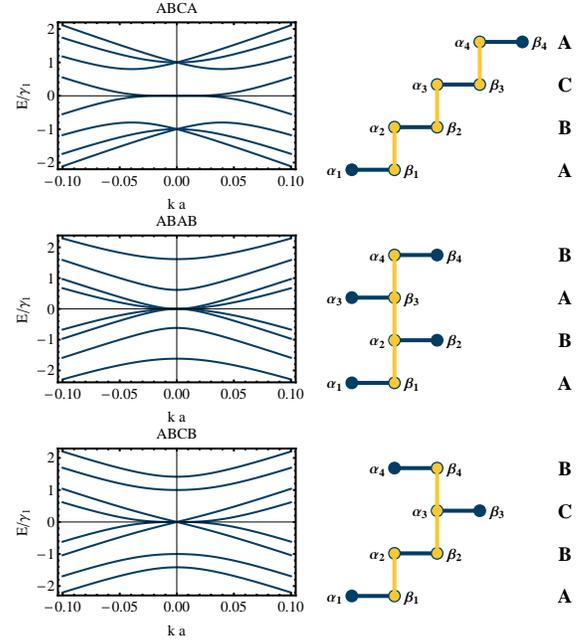}
\end{center}
\caption{(Colour online) (left) Electronic energy spectrum of the three
physically different stacking possibilities of the graphene tetralayer with only nearest neighbour interlayer hopping. The
energy is expressed in units of the interlayer hopping energy $\protect\gamma%
_1$ and the wave vector is expressed in units of $a^{-1}$, the inverse of
the nearest neighbour interatomic distance. (right) Schematic representation
of the relative positions of the sublattices $\protect\alpha$ and $\protect%
\beta$ of each layer for the stacking orders shown on the right. The yellow line corresponds to the interlayer hopping taken into account in the calculations.}
\label{QuadLayer}
\end{figure}

A system of $n$ layers of graphene can be stacked using a multitude of different stacking sequences. Graphene consists of two trigonal sublattices, called $\alpha $ and $\beta$, and it therefore suffices to consider only the relative position of these sublattices of the different layers. Due to the periodicity of the crystal, there are only three ways a layer can be placed with respect to the bottom one which we label as $A$, $B$ and $C$. It is possible to place it directly above the bottom layer (A), to shift it once (B) or twice (C) by the interatomic distance in the direction of the vector between the $\alpha $ and $\beta $ atoms. Bernal stacking (ABA)\cite{Bernal1924, Partoens2007} is found to be the most stable combination. Rhombohedral stacking (ABC) is however also possible and is observed for a small number of layers\cite{Shih2011, Craciun2009}. In between these two stacking structures, each different combination leads to a different electronic structure. To find the low energy electronic spectrum of the different possible stackings for an arbitrary number of layers, one can use the decomposition method as described by Min \textit{et al.}\cite{Min2008}. Notice however that they do not differentiate between structures that are symmetric under in plane mirroring and thus represent the same system. The total number $N\left( n\right) $ of physically different stacking possibilities for $n > 3$ layers is given by
\begin{equation}
N\left( n\right) =2^{n-2}-\sum_{i=\left\lfloor \left( n-3\right)
/2\right\rfloor }^{n-4}2^{i},
\end{equation}
where $\lfloor a\rfloor $ refers to the nearest integer smaller than $a$. The second term of this expression takes into account the mirror symmetry of a stacking sequence and is absent in the discussion of Ref.\cite{Min2008}. In Fig. \ref{QuadLayer} we show for $n=4$ the 3 stacking possibilities together with their respective energy spectrum near the $K$ point.

\section{Rhombohedral multilayers}

\label{Rhombo}

\begin{figure}[tb]
\begin{center}
\includegraphics[width=8cm]{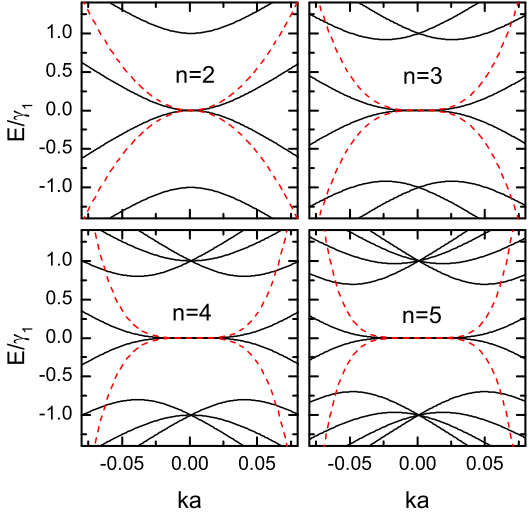}
\end{center}
\caption{(Colour online) Energy spectrum near the Dirac point of rhombohedral stacked graphene multilayers up to five layers. The black solid curves correspond to the tight binding energy with only nearest neighbour and next to nearest neighbour hopping and the red dashed curves correspond to the two band approximation. The energy is expressed in units of the interlayer hopping energy $\protect\gamma_1$ and the wave vector is expressed in units of $a^{-1}$, the inverse of the nearest neighbour interatomic distance.}
\label{Approximations}
\end{figure}

For $n$ rhombohedral (ABC) stacked graphene layers, the spectrum consists of two bands that touch at the $K$ point and $2n-2$ bands that are located at higher energies. If we only include the nearest neighbour interlayer hopping, so only hopping between the $\beta_i$ and $\alpha_{i+1}$ sublattices, the effective Hamiltonian of such system near the $K$ point is given by a $2n\times2n$ matrix
\begin{equation}
H_{n}=\hbar v_{F}\left[
\begin{array}{ccccc}
\vec{\sigma}\cdot \vec{k} & \tau & 0 & \cdots & 0 \\
\tau ^{\dag } & \vec{\sigma}\cdot \vec{k} & \tau & \cdots & 0 \\
0 & \tau ^{\dag } & \vec{\sigma}\cdot \vec{k} & \ddots & 0 \\
\vdots & \vdots & \ddots & \ddots & \tau \\
0 & 0 & 0 & \tau ^{\dag } & \vec{\sigma}\cdot \vec{k}%
\end{array}%
\right] .  \label{FullHam}
\end{equation}%
with $\vec{\sigma}=\left( \sigma _{x},\sigma _{y}\right) $ a vector of Pauli matrices, $v_{F}\approx10^6m/s$ the Fermi velocity in MLG, $\vec{k}$ the wave vector and $\tau $ is given by
\begin{equation}
\tau =\frac{1}{\hbar v_{F}}\left[
\begin{array}{cc}
0 & 0 \\
\gamma _{1} & 0%
\end{array}%
\right] ,
\end{equation}%
where $\gamma _{1}=377meV$ is the interlayer hopping parameter\cite{Partoens2006}. For this kind of stacking it is possible to introduce a two band low energy approximation which yields the Hamiltonian\cite{Min2008, Nakamura2008}
\begin{eqnarray}
H_{n}^{\prime } &=&\frac{\left( \hbar v_{F}\right) ^{n}}{\left( -\gamma
_{1}\right) ^{n-1}}\left[
\begin{array}{cc}
0 & \left( k_{x}-ik_{y}\right) ^{n} \\
\left( k_{x}+ik_{y}\right) ^{n} & 0%
\end{array}%
\right] , \\
&\sim &k^{n}\left[ \cos \left( n\phi _{k}\right) \sigma _{x}+\sin \left(
n\phi _{k}\right) \sigma _{y}\right] ,  \label{ReducedHamEq}
\end{eqnarray}%
where $\phi _{k}=\arctan \left( k_{y}/k_{x}\right) $ is the angle of the wave vector $\vec{k}$ with the normal chosen perpendicular to the pn junction and $\sigma _{x(y)}$ are the components of the pseudospin associated with this two dimensional Hamiltonian which are the respective Pauli matrices. The validity of this approximation is shown in Fig. \ref{Approximations} for $n=2,3,4,5$. In this figure we show the dispersion relation obtained by the Hamiltonian in Eq. (\ref{FullHam}) which consists of $2n$ bands. Superimposed we have plotted the dispersion relation from the Hamiltonian in Eq. (\ref{ReducedHamEq}) as dashed curves. The two band spectrum is in good agreement with the $2n$ band spectrum for low energy and near the $K$ point.

The two band Hamiltonian in Eq. $\left( \ref{ReducedHamEq}\right)$ describes a chiral particle for which the pseudospin spins $n$ times as fast as the wave vector $\vec{k}$. Notice that for the angles
\begin{equation}
\phi _{k}^{m}=m\frac{\pi }{2n} \text{ for } m\in \left\{ 1-n,\ldots ,0,\ldots
,n-1\right\} ,  \label{PhiK}
\end{equation}%
the sine or the cosine in Eq. $\left( \ref{ReducedHamEq}\right)$ vanish for $m$ even or odd respectively. At these angles the Hamiltonian commutes respectively with $\sigma _{x}$ or $\sigma _{y}$ making them conserved quantities.

When electrons impinge on a pn junction with an angle of incidence given by Eq. $\left( \ref{PhiK}\right) $, conservation of pseudospin allows the electron to be reflected only if the reflected state has the same pseudospin as the incident state. The angle of the reflected state is given by $\theta _{k}^{r}=\pi -\phi _{k}$ and therefore the wave vector of the reflected electron must rotate by an angle of $\Delta \phi =\pi -2\phi _{k}$. Since the pseudospin rotates $n$ times as fast as the wave vector, we must have $n\Delta \phi =l2\pi$, $l\in \mathbb{N}$, in order that reflection is allowed by pseudospin conservation. Using Eq. $\left( \ref{PhiK}\right) $ yields that the pseudospin of the reflected state is parallel with that of the incident state for angles $\phi _{k}^{m}$ when the difference $n-m$ is even, while the pseudospin is opposite when $n-m$ is odd.

\begin{figure}[tb]
\begin{center}
\includegraphics[width= 8cm]{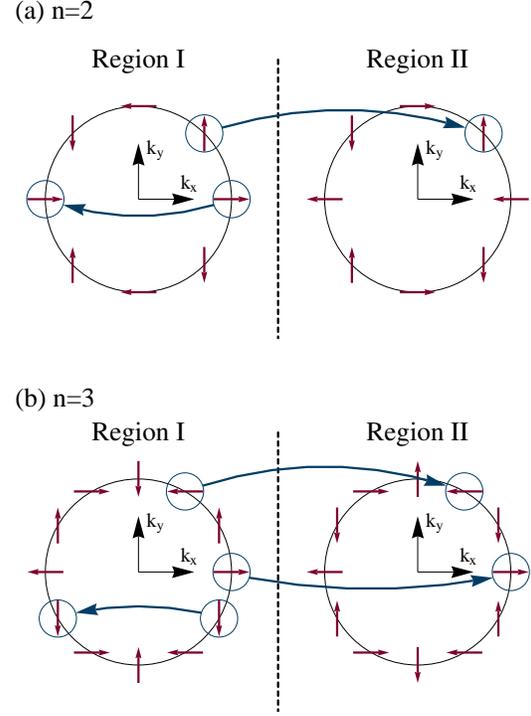}
\end{center}
\caption{(Colour online) Schematic representation of the pseudospin matching at a pn junction for (a) bilayer and (b) trilayer graphene with Fermi energy $E=V/2$. The red arrows show the rotation of the pseudospin with respect to the incident angle in the reciprocal plane outside (left) and inside (right) the junction. The blue arrows indicate the angles for which the conservation of pseudospin results in Klein tunneling (arrows directed to the right) or anti-Klein tunneling (arrows directed to the left).}
\label{Combi}
\end{figure}

When the Fermi energy $(E)$ of the incident electron is less than the potential step $(V)$ of the pn junction, the sign of the wave vector of the propagating hole state inside the junction is opposite to the sign of the incident electron wave vector. This is the result of charge conservation at the steps' edge and gives rise to a negative refraction index as found for MLG making it a meta-material\cite{Cheianov2007, RamezaniMasir2010}. Due to the change in sign of the wave vector, the angle of the wave inside the potential region also flips sign, making the pseudospin inside the junction to rotate in the opposite direction. This is schematically illustrated in Fig. \ref{Combi}. In this case the angle of refraction is given by
\begin{equation}
\theta _{k}^{t}=-\arctan \left( \frac{k_{y}}{k_{x}\left( T_{h}\right) }%
\right) ,
\end{equation}%
where $k_{x}\left( T_{h}\right) $ is the wave vector of the hole state and $T_{h}=\left\vert E-V\right\vert $ corresponds to the kinetic energy of the hole. Due to electron-hole symmetry, when $E=V/2$, the refractive angle is exactly opposite to the angle of incidence. Following a similar argument as before, one finds that for angles $\phi _{k}^{m}$ the pseudospin is opposite to that of the incident state when $n-m$ is even, but it is the same when $n-m$ is odd.

A mismatch in the pseudospin inside and outside the potential step was invoked earlier to explain Klein tunneling in MLG and anti-Klein tunneling in BLG\cite{Katsnelson2006, Gu2011}. Following the above argument, one can conclude that for an $n$ layered rhombohedral stacked system, AKT is present at angles given by Eq. $\left( \ref{PhiK}\right) $ when $n-m$ is even, while KT is present for angles given by Eq. $\left( \ref{PhiK}\right) $ when $n-m$ is odd. Table \ref{TableDing} lists the special angles for $n$ up to 5.

\begin{table}[tbp]
\caption{Angles for Klein tunneling (KT) and anti-Klein tunneling (AKT) for multilayer graphene with $n=1,\ldots,5$ layers.}
\label{TableDing}
\begin{center}
\begin{tabularx}{0.45 \textwidth}{X X X}
        \hline \hline
        n & $\phi_{KT}$ (rad) & $\phi_{AKT}$ (rad)\\
        \hline
        1 & $0$ & $-$ \\
        2 & $\pm \frac{\pi}{4}$ & $0$ \\
        3 & $0, \pm \frac{\pi}{3}$ & $\pm \frac{\pi}{6}$ \\
        4 & $\pm \frac{3\pi}{8}, \pm \frac{\pi}{8}$ & $0, \pm \frac{\pi}{4}$  \\
        5 & $0, \pm \frac{2\pi}{5}, \pm \frac{\pi}{5}$ & $\pm \frac{3\pi}{10}, \pm \frac{\pi}{10}$  \\
        \hline \hline
        \end{tabularx}
\end{center}
\end{table}

For arbitrary stacking, Min \textit{et al.}\cite{Min2008} showed that the Hamiltonian can be decomposed in a set of independent pseudospin doublets of the form
\begin{equation}
H_{J_{i}}\sim k_{i}^{J_{i}}\left[ \cos \left( J_{i}\phi _{k_{i}}\right)
\sigma _{x}+\sin \left( J_{i}\phi _{k_{i}}\right) \sigma _{y}\right] ,
\label{SubsetEq}
\end{equation}
where $k_{i}$ is the wave vector corresponding to the propagating low energy band of the $i^{th}$ pseudospin doublet and $J_{i}$ is the number of layers taking part in the doublet which corresponds to the chirality of this doublet. The number of pseudospin doublets $N_{D}$ depends on the details of the stacking, but the sum of the chiralities equals the total number of layers in the system. In this way one can decompose the low energy structure of any multilayered system in a set of $N_{D}$ non interacting doublets. Therefore, there are $N_{D}$ modes of propagation for low energy. As long as the symmetry of the system remains intact, it is not possible to scatter between different modes. The Hamiltonian given in Eq. $\left( \ref{SubsetEq}\right) $ is that of a system of $J_{i}$ rhombohedral stacked layers of graphene and because the different modes do not interact, the occurrence of KT and AKT is the same as before for a multilayer with $n=J_{i}$.

\section{Transmission probability}\label{Numerical}

\begin{figure*}[tb]
\onefigure{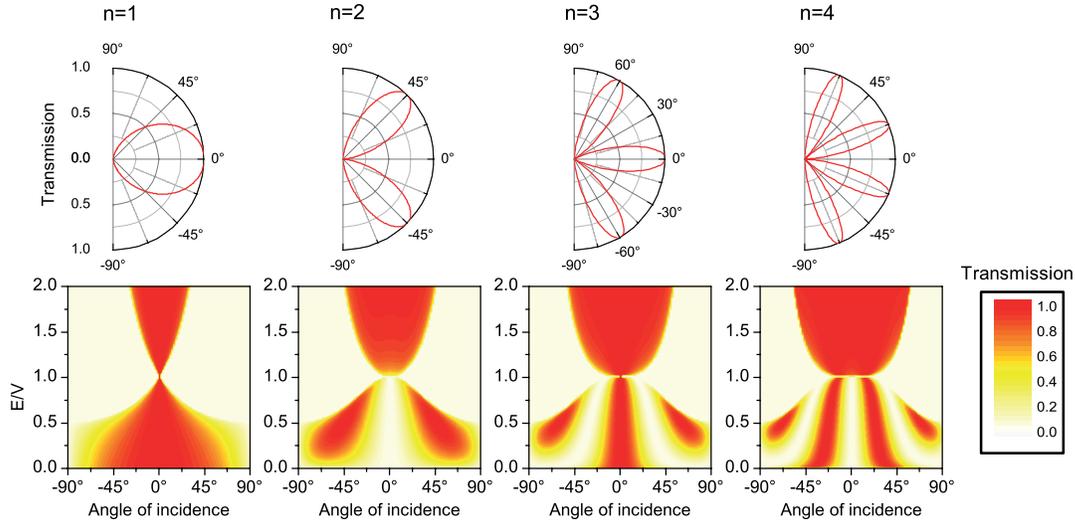}
\caption{(Colour online) Transmission probability through a pn junction in multilayer graphene for $n=1,\ldots ,4$ layers. The top row shows the angle dependend transmission for electrons with $E=V/2$. The bottom row gives countourplots of the angle and energy dependence of the transmission. }
\label{Polars}
\end{figure*}

The two band Hamiltonian in Eq. $\left( \ref{ReducedHamEq}\right) $ has a plane wave solution given by two propagating waves, one right and one left moving, and $2n-2$ evanescent waves. The wave vectors of these plane waves are the solutions of the equation
\begin{equation}
\left( k_{j}^{2}+k_{y}^{2}\right) ^{n}=\varepsilon ^{2},
\end{equation}%
where $\varepsilon =E\left( -\gamma _{1}\right) ^{n-1}/\left( \hbar v_{F}\right) ^{n}$ and $k_{y}$ is the transverse wave vector. The solution of this equation is $\pm k_{j}$ with $n$ different values for $k_{j}$. The plane wave solution can be written as a two spinor
\begin{align}
\Psi _{n}\left( x,y\right)  =&\sum_{j=1}^{n}a_{j}^{\pm }\left(
\begin{array}{c}
1 \\
\frac{\left( \pm k_{j}+ik_{y}\right) ^{n}}{\varepsilon }%
\end{array}%
\right) e^{\pm ik_{j}x+ik_{y}y} \\
=&\mathcal{PE}\left( x\right) \mathcal{C}e^{ik_{y}y},
\end{align}%
where the latter is a matrix formulation of the spinor with matrices
\begin{subequations}
\begin{align}
\mathcal{P} = & \left[
\begin{array}{cc}
1 & 1 \\
\frac{\left( k_{1}+ik_{y}\right) ^{n}}{\varepsilon } & \frac{\left(
-k_{1}+ik_{y}\right) ^{n}}{\varepsilon }%
\end{array}%
\ldots
\begin{array}{c}
 1 \\
 \frac{\left(-k_{n}+ik_{y}\right) ^{n}}{\varepsilon }
\end{array}
\right] , \\
\mathcal{E}\left( x\right)  = & Diag\left[
e^{ik_{1}x},e^{-ik_{1}x},\ldots ,e^{ik_{n}x},e^{-ik_{n}x}\right] , \\
& \text{and} & &  \notag \\
\mathcal{C} = & \left[ a_{1}^{+},a_{1}^{-},\ldots ,a_{n}^{+},a_{n}^{-}\right]
^{T}.
\end{align}%
\end{subequations}
To find the transmission probability for a pn junction, one has to equate the plane wave solutions and all the derivatives up to $n-1^{th}$ order of the region before the junction (region I) with those of the region behind it (region II) at the junction's edge at $x=0$. This leads to a set of $n$ two component equations:
\begin{equation}
\left\{
\begin{array}{c}
\mathcal{P}_{I}\mathcal{E}_{I} \mathcal{C}_{I}=\mathcal{P}%
_{II}\mathcal{E}_{II} \mathcal{C}_{II} \\
\mathcal{P}_{I}\frac{\partial \mathcal{E}_{I} }{\partial x}%
\mathcal{C}_{I}=\mathcal{P}_{II}\frac{\partial \mathcal{E}_{II}}{\partial x}\mathcal{C}_{II} \\
\vdots  \\
\mathcal{P}_{I}\frac{\partial ^{n-1}\mathcal{E}_{I} }{%
\partial x^{n-1}}\mathcal{C}_{I}=\mathcal{P}_{II}\frac{\partial ^{n-1}%
\mathcal{E}_{II}}{\partial x^{n-1}}\mathcal{C}_{II},
\end{array}%
\right.
\end{equation}%
where the matrix $\mathcal{E}$ is evaluated at $x=0$. Normalizing the incident wave on the right propagating wave before the junction by putting $a_{1,I}^{+}=1$ and applying boundary conditions $a_{j,I}^{-}=0$ and $a_{j,II}^{+}=0$ for $j\neq 1$ to suppress the non normalizable plane wave functions, the transmission $\left( T\right) $ and the reflection probability $\left( R\right) $ are given by
\begin{equation}
T=\left\vert a_{j,II}^{+}\right\vert ^{2}\text{ and }R=\left\vert
a_{j,I}^{-}\right\vert ^{2}.
\end{equation}%
The numerical results for the transmission probability for multilayers with $n=1$ up to $4$ are depicted in Fig. \ref{Polars} as function of the energy of the incident electron and the incident angle. The expected angles for KT and AKT are confirmed by our calculations.

\section{Conclusions and remarks}\label{Concl}

The combination of the chiral nature of the charge carriers and their meta-material properties induce Klein tunneling and anti-Klein tunneling at specific angles for rhombohedral stacked graphene multilayers. For an arbitrary stacking sequence, the low energy behavior of the electrons can be decomposed in independent chiral doublets with a chirality $J$ that act as if it is a rhombohedral multilayer with $n=J$ layers. In this way, for any arbitrary stacking sequence, one can predict the occurrence of Klein tunneling and anti-Klein tunneling. Note however that we limited ourselves to nearest neighbour hoppings and neglected other less important hoppings that are present in a real multilayer. The latter results in e.g. trigonal warping\cite{Koshino2009a} that will effect our results for very small energies. Furthermore, the use of the two band approximation limits the energy range to about 300meV. At high energies, additional modes of propagation need to be taken into account, changing the transmission properties for high junctions and high Fermi energy\cite{VanDuppen2012b}.

\acknowledgments
We thank S. Gillis for valuable discussions. This work was supported by the European Science Foundation (ESF) under the EUROCORES Program Euro-GRAPHENE within the project CONGRAN, and the Flemish Science Foundation (FWO-Vl).


\begin{thebibliography}{10}
\bibitem{Min2008}
\Name{Min H. \and MacDonald A.~H.} \REVIEW{Physical Review B}{77}{2008}{155416}. \\
 Please note that the $ABCB$- and the $ABAC$-tetralayer are considered to be different systems,
while it is the same under mirror symmetry and the permutation $A\rightarrow C\rightarrow B\rightarrow A$.
\end{thebibliography}
\end{document}